\pdfoutput=1
\documentclass[journal]{IEEEtran}
\usepackage{cite}
\usepackage{amsmath,amssymb,amsfonts}
\usepackage{graphicx}
\usepackage{booktabs}
\usepackage{algorithm}
\usepackage{algorithmic}
\usepackage{multirow}
\usepackage{array}
\usepackage{url}
\usepackage[hidelinks]{hyperref}
\usepackage{orcidlink}

\begin{document}
\bstctlcite{BSTcontrol}

\title{TIRMamba: A Thermal-Prior-Modulated State-Space Network\\
for Sub-Million-Parameter Infrared Image Super-Resolution}

\author{Chun-An~Lin\,\orcidlink{0009-0004-8874-6196},~\IEEEmembership{Student Member,~IEEE,}
        Tsung-Jung~Liu\,\orcidlink{0000-0003-4296-0942},~\IEEEmembership{Member,~IEEE,}
        and~Yen-Chieh~Ouyang\,\orcidlink{0000-0002-4221-2787},~\IEEEmembership{Life Member,~IEEE}%
\thanks{This work has been submitted to the IEEE for possible publication. Copyright may be transferred without notice, after which this version may no longer be accessible.}%
\thanks{This work was supported in part by the National Chung-Shan Institute of Science and Technology, Taiwan, and in part by the National Science and Technology Council, Taiwan, under Grant NSTC 114-2221-E-005-040-MY2. \textit{(Corresponding author: Tsung-Jung Liu.)}}%
\thanks{C.-A. Lin is with the Department of Electrical Engineering, National Chung Hsing University, Taichung 402202, Taiwan, and also with the College of Computing and Informatics, Providence University, Taichung 433303, Taiwan (e-mail: julian135707@gmail.com).}%
\thanks{T.-J. Liu and Y.-C. Ouyang are with the Department of Electrical Engineering, National Chung Hsing University, Taichung 402202, Taiwan (e-mail: tjliu@email.nchu.edu.tw; ycouyang@nchu.edu.tw).}%
}

\markboth{Submitted to IEEE Transactions on Geoscience and Remote Sensing}%
{Lin \MakeLowercase{\textit{et al.}}: TIRMamba for Infrared Image Super-Resolution}

\maketitle

\begin{abstract}
Infrared image super-resolution is currently led by Mamba-based networks with 26 to 37 million parameters, which are difficult to deploy on the airborne and handheld platforms where thermal imaging is most needed. This paper presents TIRMamba, a network with 896K to 910K parameters for single-channel thermal imagery. A Thermal Prior Highway computes gradient, local-contrast and spectral cues once at the input and, through one adapter per residual group, modulates a weight-tied bidirectional state-space trunk and gates its dual-scale detail branch; a tri-path reconstruction adds the learned residual to a bicubic radiometric baseline. Because the standard benchmark provides only 265 infrared training images and evaluates fusion products on full images, we train with a replay strategy: grayscale DIV2K pre-training followed by fine-tuning on 64-pixel patches drawn with equal probability from the infrared and natural corpora. At scale factor 4, TIRMamba matches the strongest protocol-trained methods on both official test sets with 29 to 40 times fewer parameters and 2.8 to 9.4 times lower latency; at scale factor 2 it gives the highest SSIM on both. A variant with prior-conditioned selectivity, TIRMamba-Rad, corrects a 3\,dB raw-thermal failure of an intermediate size-invariant design and gives the best results at scale factor 4 on raw-thermal, unmanned-aerial-vehicle and independent-sensor test sets. Code and trained models will be released at \url{https://github.com/julian135707/TIRMamba} upon acceptance.
\end{abstract}


\section{Introduction}\label{sec:intro}
\IEEEPARstart{T}{hermal} infrared (IR) cameras sense emitted radiation rather than reflected light, so they remain effective in darkness, smoke, haze, and glare. On unmanned aerial vehicles (UAVs), aircraft and ground platforms, thermal imaging supports search and rescue, wildfire and disaster response, environmental monitoring, and combined thermal--visible perception~\cite{liu2022tardal}. Its main weakness is spatial resolution: long wavelengths set a coarse diffraction limit, microbolometer pixel pitch is difficult to reduce, and high-definition focal-plane arrays are expensive, so single-image super-resolution (SISR) is an attractive alternative to optical upgrades on weight- and power-constrained platforms~\cite{huang2025irsurvey}.

\begin{figure}[!t]
\centering
\includegraphics[width=\columnwidth]{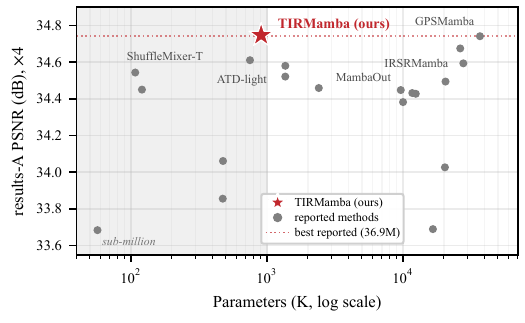}
\caption{Accuracy versus model size for $\times4$ super-resolution on results-A. Gray points are the methods of the source benchmark at their reported scores (Table~\ref{tab:x2} and Table~S-VIII of the supplementary material; the anomalous MambaIRv2 entry is omitted), and the shaded band marks the sub-million-parameter regime. TIRMamba matches the best reported score with $29$ to $40$ times fewer parameters.}
\label{fig:scatter}
\end{figure}

Deep SISR has moved from convolutional and adversarial networks through window-attention transformers to selective state-space models (SSMs), such as Mamba~\cite{gu2023mamba}, whose global receptive field at linear cost has proved effective for restoration~\cite{guo2024mambair,guo2025mambairv2,li2025mair}. Infrared SR has followed the same path, and the Mamba-based IRSRMamba~\cite{huang2025irsrmamba} and GPSMamba~\cite{huang2025gpsmamba} now lead the standard benchmark with $26.5$M and $36.9$M parameters. Models of this size are ill suited to the settings in which infrared SR is most needed: UAVs, handheld scopes and vehicle-embedded thermal modules have far less processing power, memory and energy than a workstation GPU. Efficient SR architectures exist~\cite{sun2023shufflemixer,sun2023safmn,liu2025catanet,zhang2024atd}, lightweight design is an active topic in remote sensing~\cite{sui2023gcrdn,hua2024rcln}, and attention-based lightweight networks have been built for infrared imagery specifically~\cite{lin2024lirsrn}, but on this benchmark sub-million-parameter models have so far scored below the large designs (Fig.~\ref{fig:scatter}).

Two properties of the benchmark protocol~\cite{huang2021psrgan,huang2025irsrmamba} shape any lightweight solution. The protocol provides only $265$ infrared training images, and its two test sets are infrared--visible fusion products that retain visible-band texture. Networks trained only on natural images therefore already reach the published state of the art, whereas fine-tuning on the $265$ images alone moves a model away from the test statistics (Sec.~\ref{ssec:protocol_effects}). The protocol also trains on crops of $8$ to $16$ pixels but evaluates full images, which leaves global operators such as a state-space scan or a Fourier layer poorly calibrated at test time~\cite{chu2022tlc}.

This paper presents TIRMamba, a sub-million-parameter SR network for single-channel thermal imagery, designed together with a training strategy that addresses both effects (Sec.~\ref{sec:method}, Fig.~\ref{fig:arch}). Physically motivated infrared cues are computed once at the input and carried through a lightweight Thermal Prior Highway (TPH) that modulates a weight-tied bidirectional state-space trunk, and a dual-scale detail branch~\cite{chen2022nafnet,guo2024mambair} supplies the local structure that a scan alone forgets. 

The contributions of this paper are as follows.
\begin{itemize}
\item \textbf{Prior-modulated state-space backbone.} TIRMamba is, to our knowledge, the first sub-million-parameter state-space network for single-channel infrared SR on this benchmark ($896$K to $910$K parameters). Its Thermal Prior Highway carries gradient, local-contrast and spectral cues computed once at the input, and one zero-initialized adapter per residual group turns them into a spatially varying FiLM modulation of the scanned input and a gate on a dual-scale detail branch. A tri-path reconstruction adds the learned residual to a bicubic radiometric baseline (Sec.~\ref{sec:method}).
\item \textbf{Prior-conditioned selectivity.} TIRMamba-Rad adds the thermal prior to the step size $\Delta$ of the selective scan through a zero-initialized projection and anchors the DC level to the bicubic term (Sec.~\ref{ssec:plus}); to our knowledge it is the first infrared SR network to tie the scan timescale to an externally computed prior. It removes a $3$\,dB raw-thermal failure of an intermediate size-invariant design, a failure the fusion benchmark cannot detect, and gives the best $\times4$ results on the raw-thermal, UAV and independent-sensor test sets.
\item \textbf{Protocol-grounded training strategy.} Two measured properties of the protocol motivate the strategy: its fusion-derived test sets reward visible-band statistics, and its $8$- to $16$-pixel training crops miscalibrate global operators on full images. Grayscale DIV2K pre-training~\cite{agustsson2017div2k} followed by replay-mixed fine-tuning~\cite{rolnick2019experience} on $64$-pixel patches drawn equally from M3FD-265 and DIV2K addresses both (Sec.~\ref{ssec:training}); controls isolate each effect, and the strategy transfers to a second lightweight backbone, ATD-light (Sec.~\ref{ssec:sota}).
\item \textbf{Evaluation.} At $\times4$, TIRMamba matches IRSRMamba and GPSMamba on both official test sets with $29$ to $40$ times fewer parameters and $2.8$ to $9.4$ times lower latency (Fig.~\ref{fig:scatter}), and at $\times2$ it gives the highest SSIM on both while running $9.9$ to $44$ times faster. Per-image statistical tests, three additional thermal test sets and a strict-protocol control support these claims (Secs.~\ref{sec:experiments} and~\ref{sec:generalization}).
\end{itemize}

\section{Related Work}\label{sec:related}
\subsection{Lightweight Single-Image Super-Resolution}
Deep SISR developed from convolutional networks~\cite{dong2015srcnn,dong2016fsrcnn,lim2017edsr,zhang2018rcan} through adversarial training~\cite{ledig2017srgan,wang2018esrgan} to window-attention transformers~\cite{liang2021swinir,chen2023hat,chen2024rgt,zhang2024atd}, and the current lightweight designs replace dense attention with cheaper token mixing, spatially adaptive modulation or content-aware aggregation~\cite{sun2023shufflemixer,sun2023safmn,liu2025catanet}; the NTIRE efficient SR challenge ranks such models by runtime, parameter count and FLOPs under a PSNR floor~\cite{ren2025ntire}. TIRMamba reuses the simplified channel attention and SimpleGate of NAFNet~\cite{chen2022nafnet} and applies the pointwise frequency-domain filter of SwinFIR~\cite{zhang2023swinfir} once at the input rather than in every block. Only ShuffleMixer, CATANet and the $753$K ATD-light have protocol-trained infrared scores (Table~\ref{tab:x2}), all below the $26.5$M and $36.9$M leaders, and none was designed for thermal input.

\subsection{State-Space Models for Vision and Image Restoration}
Structured state-space models offer a global receptive field at linear cost, and Mamba~\cite{gu2023mamba} made the state transition input-dependent, or \emph{selective}. Vision Mamba (Vim, listed as VisionMamba in Table~\ref{tab:x2})~\cite{zhu2024vim} applies bidirectional selective-scan blocks to a position-embedded patch sequence, and its speed and memory advantages over attention grow with resolution; VMamba~\cite{liu2024vmamba} instead runs four cross-scan routes and merges them. Han \emph{et al.}~\cite{han2024demystify} view the selective SSM as a variant of linear attention and trace its advantage in vision mainly to the forget gate and the block design rather than to the scan itself. MambaOut~\cite{yu2025mambaout} argues that SSM token mixing is warranted only for tasks with long sequences or autoregressive structure. ImageNet classification has neither, and gated-CNN blocks without the SSM surpass every visual Mamba model on that task but fall short on detection and segmentation, which the authors take as evidence that the SSM remains worthwhile for long-sequence visual tasks. Whole-image SR is such a task, at $19.6$K to $110.6$K tokens per $\times2$ test image, and on it the SR adaptations of the Vim and MambaOut backbones reported by the source benchmark trail the two infrared-specific designs, IRSRMamba and GPSMamba (Table~\ref{tab:x2}). 

MambaIR~\cite{guo2024mambair} counters local pixel forgetting and channel redundancy with local convolution and channel attention; MambaIRv2~\cite{guo2025mambairv2} shows that a single attentive scan can replace four-direction scanning; and MaIR~\cite{li2025mair} preserves locality with nested S-shaped stripe scans at no added cost. Lightweight Mamba SR has been built for RGB imagery, from a $0.424$M distilled model to transformer-hybrid designs~\cite{lei2024dvmsr,guo2026tpmambasr,zhu2025fssm}. The MambaIR, MambaIRv2 and MaIR findings guide three choices in TIRMamba: stripe-serpentine scanning, shared bidirectional SSM weights, and identity-initialized channel recalibration. What TIRMamba adds is modulation of the scanned input by a thermal prior, the per-channel scale and shift of FiLM~\cite{perez2018film} made spatially varying as in SFT~\cite{wang2018sft}, rather than the learned prompt components of PromptIR~\cite{potlapalli2023promptir} or the semantic-frequency prompt of GPSMamba~\cite{huang2025gpsmamba}. TIRMamba-Rad further conditions the step size $\Delta$ on that prior; all-in-one restoration has modulated $\Delta$ with degradation-aware signals~\cite{gao2026multitau}, but to our knowledge no infrared SR network has tied it to an externally computed physical prior.

\subsection{Infrared and Remote-Sensing Super-Resolution}
Thermal images have a limited gradient range~\cite{qin2024lkformer} and low pattern diversity~\cite{huang2021hetsrwgan}, and their resolution is capped by optics, fixed-pattern noise and on-board processing~\cite{huang2025irsurvey}. Two benchmark families exist. The PBVS thermal SR challenge series uses a multi-camera thermal dataset, and its 2025 edition ran single-image $\times8$ and RGB-guided tracks led by transformer and hybrid entries~\cite{rivadeneira2025pbvs}. The second family, used here, is the results-A/results-C suite introduced with PSRGAN~\cite{huang2021psrgan}, a two-path progressive GAN with multistage visible-to-infrared transfer. LKFormer, CoRPLE and DifIISR~\cite{qin2024lkformer,huang2024corple,li2025difiisr} report on other M3FD splits or their own data, and lightweight infrared designs rely on convolution~\cite{gulisn2024}, adversarial training~\cite{huang2021hetsrwgan} or multi-domain attention~\cite{lin2024infrared}. IRSRMamba~\cite{huang2025irsrmamba} adds wavelet feature modulation and a semantic-consistency loss to a Mamba backbone, and GPSMamba~\cite{huang2025gpsmamba} adds a fused semantic-frequency prompt to the output matrix $C$ of the selective scan in each block under a phase-consistency loss; they lead the benchmark at $26.5$M and $36.9$M reported parameters. Physical cues appear elsewhere too: SCINet~\cite{fang2024scinet} feeds local-contrast and gradient priors through a contrast branch, and DifIISR applies Fourier-magnitude guidance at each sampling step; TIRMamba computes such cues once and routes them through one conditioning stream.

Remote-sensing SR has followed the same trajectory as generic SISR, from lightweight CNNs~\cite{wang2022fenet,sui2023gcrdn,hua2024rcln} and the token-selective transformer TTST~\cite{xiao2024ttst} to state-space designs: RS-Mamba~\cite{zhao2024rsmamba} applies a selective scan to whole very-high-resolution images for dense prediction, and Mamba-based remote-sensing SR adds frequency selection~\cite{xiao2025fmsr} and cross-scale state propagation~\cite{jiang2025repmamba}. Aerial infrared SR has become a remote-sensing task in its own right: the top three entries of the NTIRE 2026 remote-sensing infrared SR challenge all build on HAT~\cite{chen2023hat,liu2026ntire}. None of the state-space designs above targets single-channel thermal imagery, and to our knowledge no sub-million-parameter state-space network for such input has been reported on the PSRGAN/IRSRMamba protocol; we follow that protocol (Sec.~\ref{ssec:datasets}) and quote reported scores unchanged.

\section{Proposed Method: TIRMamba}\label{sec:method}

\subsection{Overview}
\label{ssec:overview}

\begin{figure*}[!t]
\centering
\includegraphics[width=0.95\textwidth]{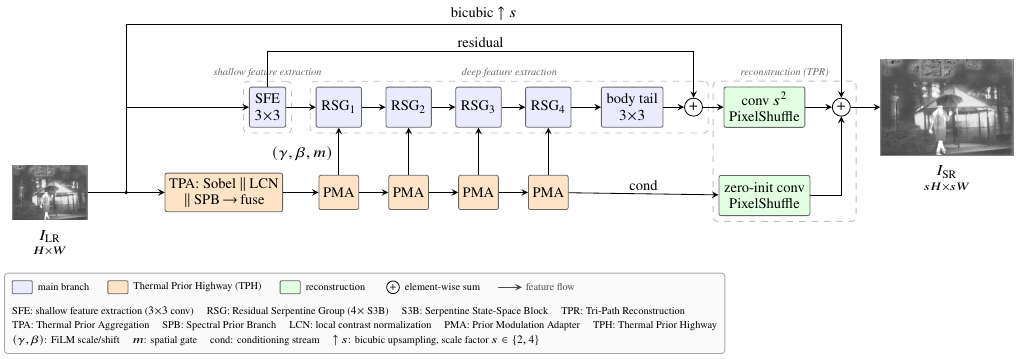}
\caption{Architecture of TIRMamba. Colors group the main branch, the Thermal Prior Highway and the reconstruction stage; symbols and abbreviations are defined in the legend. The same network serves both scale factors $s\in\{2,4\}$; the input--output pair is a $\times4$ example from results-C.}
\label{fig:arch}
\end{figure*}

TIRMamba takes a single-channel low-resolution image $x\in\mathbb{R}^{H\times W}$ and returns $I_{\mathrm{SR}}\in\mathbb{R}^{sH\times sW}$ for a scale factor $s\in\{2,4\}$, in the three stages common to SR networks (Fig.~\ref{fig:arch}). The shallow feature extraction (SFE) stage is one $3{\times}3$ convolution that maps $x$ to $64$ channels; in parallel, a \emph{Thermal Prior Aggregation} (TPA) module computes infrared cues from $x$ and starts a conditioning stream, the \emph{Thermal Prior Highway} (TPH), that runs alongside the trunk. The deep feature extraction stage contains four \emph{Residual Serpentine Groups} (RSGs), each of four \emph{Serpentine State-Space Blocks} (S3Bs) followed by a depthwise-separable tail and a group residual, and is closed by a $3{\times}3$ body tail and a global residual. Before each group a \emph{Prior Modulation Adapter} (PMA) advances the TPH by one step and produces the modulation signals that the four blocks share. A \emph{Tri-Path Reconstruction} (TPR) stage maps the trunk output $F$ and the final TPH state $c_4$ back to image space.

Two principles guide the design. Every infrared-specific mechanism enters through a side path whose last layer is initialized at zero, so at initialization the network is a generic residual Mamba SR backbone and the priors are learned only where they reduce the loss. Parameters are saved by structure rather than by narrowing the width: the two scan directions share one SSM, one adapter serves a whole group, group tails are depthwise-separable, and upsampling is a single step, which gives $896.1$K parameters at $\times2$, $910.0$K at $\times4$, and $16.27$\,GFLOPs for a $135\times180$ input. The network processes the single sensor channel throughout (Sec.~\ref{ssec:datasets}).

\subsection{Thermal Prior Highway (TPH)}
\label{ssec:prior}

\subsubsection{Thermal Prior Aggregation}
\begin{figure*}[!t]
\centering
\includegraphics[width=0.80\textwidth]{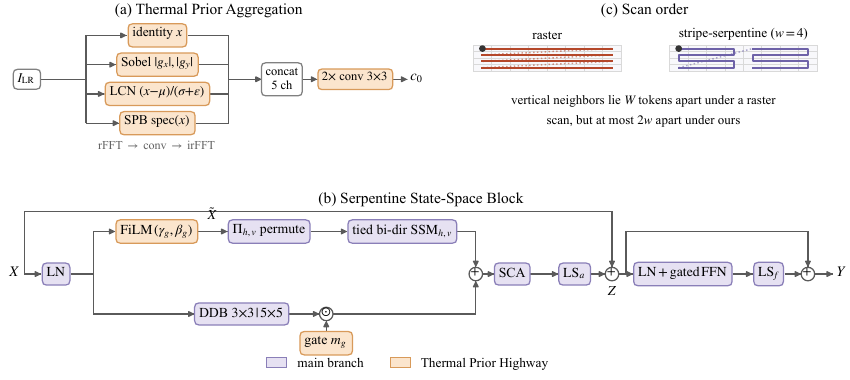}
\caption{Module details, with the same color coding as Fig.~\ref{fig:arch}.
(a) Thermal Prior Aggregation: the gradient, local-contrast and spectral cues are
concatenated with the input and fused into the conditioning stream $c_0$.
(b) Serpentine State-Space Block, drawn to match (\ref{eq:s3b}). (c) Scan order, drawn with $w=4$ for clarity (the network uses $w=8$). Raster flattening separates vertically adjacent
pixels by one image width, whereas the stripe-serpentine order keeps them
within $2w$ sequence positions.}
\label{fig:modules}
\end{figure*}

Three properties of thermal imagery can be measured directly from the input: boundaries are weak and wide, radiometric contrast is locally compressed, and sensor structure such as fixed-pattern noise occupies narrow bands of spatial frequency. TPA computes one cue for each, once and at full resolution (Fig.~\ref{fig:modules}(a)); the cues themselves are generic operators, and what is specific to thermal imagery is the choice of cues and the gated path through which they enter the trunk.

\emph{Gradient cues.} $|g_x|$ and $|g_y|$ are the absolute responses of the horizontal and vertical Sobel filters, with the $3{\times}3$ kernels divided by $4$ so that the response has the same range as $x$. They locate weak edges.

\emph{Local contrast normalization.} $\mathrm{LCN}(x)=(x-\mu)/(\sigma+\varepsilon)$, where $\mu$ and $\sigma$ are the mean and standard deviation of $x$ over a $5{\times}5$ window centered on each pixel, computed by average pooling, and $\varepsilon=10^{-6}$ prevents division by zero on flat regions. LCN has no parameters and amplifies low-contrast targets regardless of their absolute radiometric level.

\emph{Spectral Prior Branch (SPB).} We apply the two-dimensional discrete Fourier transform for real input (rFFT2) to $x$ with orthonormal scaling. The result is a complex array; we stack its real and imaginary parts as two channel groups, pass them through two $1{\times}1$ convolutions with a LeakyReLU (slope $0.1$) between them, split the output back into a real and an imaginary group, and invert the transform. A $1{\times}1$ convolution after the inverse transform gives the $32$-channel spatial map $\mathrm{spec}(x)$. A pointwise operation in the frequency domain is a learnable global filter~\cite{zhang2023swinfir}, so every pixel of $\mathrm{spec}(x)$ has an image-wide receptive field and the branch can address artifacts concentrated at particular frequencies; spatial--frequency attention of this kind has also proved useful in lightweight infrared SR~\cite{lin2025lightweight}.

The five maps $x$, $|g_x|$, $|g_y|$, $\mathrm{LCN}(x)$ and $\mathrm{spec}(x)$ are concatenated and fused by a $3{\times}3$ convolution, a LeakyReLU (slope $0.1$) and a second $3{\times}3$ convolution into the $64$-channel initial state $c_0$ of the TPH.

\subsubsection{Prior Modulation Adapter}
Before group $g\in\{1,\dots,4\}$, a PMA updates the stream with a residual step,
\begin{equation}
c_g = c_{g-1} + W_p\,\phi\!\left(W_d\,c_{g-1}\right),
\label{eq:pma}
\end{equation}
where $W_d$ is a depthwise $3{\times}3$ convolution, $\phi$ is a LeakyReLU with slope $0.1$, and $W_p$ is a $1{\times}1$ convolution whose weights and bias are initialized at zero, so that $c_g=c_{g-1}$ at initialization. From $c_g$ the adapter emits two signals shared by the four blocks of the group. The first is a pair of FiLM~\cite{perez2018film} maps $(\gamma_g,\beta_g)=W_f\,c_g$, with $W_f$ a zero-initialized $1{\times}1$ convolution, applied to the normalized block input as $(1+\gamma_g)\odot\mathrm{LN}(X)+\beta_g$; the identity is recovered at initialization. The second is a spatial gate
\begin{equation}
m_g=\operatorname{sigm}\!\left(W_m\,c_g\right),
\label{eq:gate}
\end{equation}
where $\operatorname{sigm}(\cdot)$ is the logistic function (we reserve $\sigma$ for the local standard deviation in LCN) and $W_m$ is a $1{\times}1$ convolution with zero weights and bias $2$, so $m_g\approx0.88$ everywhere at initialization and the gate starts close to pass-through. This is the spatially varying modulation of SFT~\cite{wang2018sft} applied to a state-space trunk. Because Mamba derives its dynamics from the scanned input itself~\cite{gu2023mamba}, modulating that input with edge, contrast and spectral evidence guides state selection without touching the SSM kernel, whereas the frequency prompt of GPSMamba~\cite{huang2025gpsmamba} is injected inside it. One adapter per group instead of one per block divides the adapter parameters by about four.

\subsection{Serpentine State-Space Block (S3B)}
\label{ssec:block}

\subsubsection{Stripe-serpentine serialization}
A state-space model consumes a one-dimensional sequence, and raster flattening places vertically adjacent pixels $W$ positions apart at every line wrap. Following the stripe serialization of MaIR~\cite{li2025mair}, we divide the image into vertical stripes of width $w=8$ and traverse each stripe row by row in serpentine order, reversing direction between rows and between stripes. Vertically adjacent pixels inside a stripe are then at most $2w$ positions apart, and the jump between stripes is at most $w{+}1$ pixels (Fig.~\ref{fig:modules}(c)). The serialization is an index permutation $\Pi$, adds no parameters or FLOPs, and is cached per resolution.  We adopt this order for the locality argument; our controlled test does not show an accuracy gain over raster order at this model size (Sec.~\ref{ssec:module_ablation}).

\subsubsection{Half-channel directional split}
A $1{\times}1$ projection precedes the scan. Half of the channels are then serialized with vertical stripes ($\Pi_h$, horizontal adjacency at unit distance) and the other half with the transposed pattern ($\Pi_v$, vertical adjacency at unit distance). Any pair of adjacent pixels is within about $2w$ positions in at least one branch, which keeps the benefit of multi-directional scanning without running every direction on all channels, a redundancy reported in~\cite{guo2025mambairv2}.

\subsubsection{Tied bidirectional SSM}
Each branch is scanned in both directions by the same SSM. The reversed sequence is stacked along the batch dimension, processed in the same kernel call, reversed back and averaged with the forward output. The reversal of a serpentine path is itself a serpentine path, so the shared weights see statistically equivalent inputs, and tying halves both the directional SSM parameters and the number of kernel launches. Each branch is a Mamba-1 selective SSM~\cite{gu2023mamba} of width $32$, state dimension $16$ and expansion factor $1$; we write $\mathrm{SSM}_{h,v}$ for the two tied bidirectional branches together with the concatenation of their outputs.

\subsubsection{Detail branch, channel attention and feed-forward network}
SSMs used alone suffer from local pixel forgetting and channel redundancy~\cite{guo2024mambair}. A parallel Dual-Scale Detail Branch (DDB) of depthwise convolutions, $3{\times}3$ on one half of the channels and $5{\times}5$ on the other to match the wide edge transitions of thermal images, supplies the local term; the gate $m_g$ scales it and can suppress it over flat backgrounds. The two streams are summed and recalibrated by the simplified channel attention (SCA) of NAFNet~\cite{chen2022nafnet}, global average pooling followed by one $1{\times}1$ convolution initialized to the identity, and a $1{\times}1$ output projection completes the block. Each block then applies a gated feed-forward network (FFN): a $1{\times}1$ convolution expands the width by $1.5$ and doubles it again for the gate, a depthwise $3{\times}3$ convolution follows, the two halves are multiplied (SimpleGate~\cite{chen2022nafnet}), and a $1{\times}1$ convolution projects back to $64$ channels.

With $\mathrm{LN}$ a per-pixel layer normalization over channels, $\odot$ the element-wise product, and $\mathrm{LS}_a,\mathrm{LS}_f\in\mathbb{R}^{64}$ per-channel LayerScale vectors~\cite{touvron2021cait} initialized at $10^{-2}$, one block maps its input $X$ to its output $Y$ as
\begin{equation}
\begin{aligned}
\tilde{X} &= (1+\gamma_g)\odot\mathrm{LN}(X)+\beta_g,\\
Z &= X + \mathrm{LS}_a\odot\mathrm{SCA}\bigl(\mathrm{SSM}_{h,v}(\Pi_{h,v}(\tilde{X}))\\
&\qquad\qquad\quad + m_g\odot\mathrm{DDB}(\mathrm{LN}(X))\bigr),\\
Y &= Z + \mathrm{LS}_f\odot\mathrm{FFN}\!\left(\mathrm{LN}(Z)\right),
\end{aligned}
\label{eq:s3b}
\end{equation}
where the FiLM signal modulates only the scanned input, the detail branch reads the normalized input directly, and the input and output $1{\times}1$ projections of the block are absorbed into $\mathrm{SSM}_{h,v}$ and $\mathrm{SCA}$ (Fig.~\ref{fig:modules}(b)).

\subsection{Tri-Path Reconstruction (TPR)}
\label{ssec:recon}

The reconstruction stage sums three paths,
\begin{equation}
I_{\mathrm{SR}} = \mathrm{bicubic}_{\times s}(x) + R(F) + H(c_4),
\label{eq:tpr}
\end{equation}
where $F$ is the trunk output and $c_4$ is the final TPH state. The main path $R$ is one $3{\times}3$ convolution from $64$ to $s^2$ channels followed by PixelShuffle~\cite{shi2016pixelshuffle}. The usual full-width head, a $3{\times}3$ convolution from $64$ to $64s^2$ channels, PixelShuffle, and a $3{\times}3$ convolution from $64$ to $1$ channel, has $64\cdot64\cdot4\cdot9=147{,}456$ weights at $s=2$ plus biases and the final layer, about $148$K; $R$ has $64\cdot4\cdot9+4=2{,}308$, a saving of about $145$K parameters. The high-frequency skip $H$ is a zero-initialized $3{\times}3$ convolution to $s^2$ channels followed by PixelShuffle, applied to $c_4$, so edge and spectral evidence reaches the output directly without disturbing early training. The bicubic term carries the radiometric level, so the two learned paths predict only residual detail, a bias suited to thermal imagery, in which the absolute level has physical meaning.

\subsection{TIRMamba-Rad and Prior-Conditioned Selectivity}
\label{ssec:plus}

TIRMamba-Rad ($948$K parameters at $\times2$, $990$K at $\times4$) is the same backbone tuned for unfused thermal radiometry, where large near-zero backgrounds make the absolute background level, rather than texture, the main source of error. It was motivated by a $3$\,dB failure on raw thermal data of an intermediate configuration, a $912.9$K flag-controlled superset of TIRMamba described at the end of this subsection, that the fusion test sets cannot show and that Sec.~\ref{ssec:secondgen} traces to its spectral prior. TIRMamba-Rad keeps six of the seven refinements of that configuration, all but the scale-invariant spectral prior, raises the $\Delta$ projection rank to $16$, widens the detail branch to all channels with three scales ($3{\times}3$, $5{\times}5$ and a dilated $3{\times}3$), and makes five further changes; Table~S-IV of the supplementary material lists the complete configuration of the three models side by side.

\emph{No SPB.} The spectral cue is removed from TPA, whose input becomes $[x,|g_x|,|g_y|,\mathrm{LCN}(x)]$, and the state-space path takes over its global-context role.

\emph{DC-anchored reconstruction.} We call the spatial mean of an image its DC level, that is, its zero-frequency component or absolute radiometric level. With $\operatorname{zeroDC}(r)=r-\operatorname{mean}_{h,w}(r)$, the reconstruction becomes
\begin{equation}
\begin{aligned}
I_{\mathrm{SR}} = \mathrm{bicubic}_{\times s}(x) + D(F) &+ \operatorname{zeroDC}\bigl(R(F)\bigr)\\ &+ \operatorname{zeroDC}\bigl(H(c_4)\bigr),
\end{aligned}
\label{eq:tprdc}
\end{equation}
where $D$ is a $3{\times}3$ convolution to $s^2$ channels followed by PixelShuffle, initialized with weights of standard deviation $10^{-3}$ and zero bias. $D$ is the only learned path that can change the DC level; the other two contribute high-frequency detail only, so background levels start from the bicubic anchor.

\emph{$\gamma$-only FiLM.} The adapter emits $\gamma_g$ alone and the modulated input becomes $\tilde{X}=(1+\gamma_g)\odot\mathrm{LN}(X)$; the additive term $\beta_g$ was the remaining route by which the TPH could inject a constant offset into the trunk.

\emph{Prior-conditioned selectivity (PCS).} In a selective SSM the step size $\Delta$ is computed from the scanned tokens, so the timescale of the recurrence depends on the trunk features alone; PCS adds the TPH to that computation. For token $n$ of a branch, with $u_n$ the low-rank projection of the token that Mamba uses for $\Delta$ and $c_g[n]$ the TPH value at the same pixel under the same permutation,
\begin{equation}
\Delta_n=\operatorname{softplus}\!\left(W_\Delta u_n + b_\Delta + W_\delta\, c_g[n]\right),
\label{eq:pcs}
\end{equation}
where $W_\Delta,b_\Delta$ are the standard Mamba projection and $W_\delta$ is a zero-initialized $1{\times}1$ convolution produced by the adapter, so PCS is inert at initialization and $\Delta_n$ stays positive. Edge, contrast and radiometric cues can then shorten or lengthen the memory of the scan directly; Sec.~\ref{sec:related} places this next to the degradation-conditioned step sizes of recent restoration work and the output-side prompt of GPSMamba. 

\emph{First-order hold.} The zero-order-hold discretization of Mamba is replaced by a first-order hold, in which the input to the recurrence at step $n$ is $u_n=(x_n+x_{n+1})/2$; this has no parameters and was recently shown to help lightweight SR state-space models~\cite{zhu2025fssm}.

TIRMamba-Rad is a second operating point rather than a replacement: it scores below the base model on the $\times4$ fusion sets and above it on the test sets of Sec.~\ref{sec:generalization}, so the fusion comparisons use the base model, and the remaining test sets are reported for both. The $912.9$K intermediate configuration, a flag-controlled superset of TIRMamba trained at $\times2$ only, is the base of the component ablation (Table~\ref{tab:ablation_module}); its forward-normalized SPB with $\operatorname{asinh}$ compression is the size-invariance route of Sec.~\ref{ssec:protocol_effects} and the component removed above.

\subsection{Training Strategy}
\label{ssec:training}

The benchmark permits $265$ infrared training images, and Sec.~\ref{ssec:protocol_effects} measures two costs of that protocol: adapting on $265$ images lowers results-C whichever modality the images come from, and fine-tuning on $16$-pixel crops leaves the global operators mismatched to full-image inference. The training strategy addresses both. We first pre-train on grayscale sub-images of DIV2K~\cite{agustsson2017div2k}. We then fine-tune with experience replay~\cite{rolnick2019experience}: each training patch is drawn with probability $0.5$ from M3FD-265 and $0.5$ from the grayscale DIV2K pool, so the optimizer stays within the support of the pre-training distribution while learning infrared statistics. Fine-tuning uses low-resolution patches of $64$ pixels at $\times2$ and $32$ pixels at $\times4$ ($128$ HR pixels in both cases), which reduces the gap between training and test statistics of the scan and of the spectral branch. The loss is
\begin{equation}
\mathcal{L}=0.9\,\mathcal{L}_{\mathrm{Char}}+0.1\,\bigl\lVert\mathcal{F}(I_{\mathrm{SR}})-\mathcal{F}(I_{\mathrm{HR}})\bigr\rVert_{1},
\label{eq:loss}
\end{equation}
where $\mathcal{L}_{\mathrm{Char}}=\operatorname{mean}\sqrt{(I_{\mathrm{SR}}-I_{\mathrm{HR}})^{2}+\epsilon}$ is the Charbonnier penalty~\cite{lai2017lapsrn} with $\epsilon=10^{-6}$, $\mathcal{F}$ is the two-dimensional discrete Fourier transform of the image, and the second term is the mean absolute value of the complex difference over all frequency bins, a frequency-domain term as used in~\cite{sun2023safmn}. The $\times4$ model is initialized from the best $\times2$ replay checkpoint with a higher learning rate on its scale-specific reconstruction head; Sec.~\ref{ssec:implementation} lists the optimizer, schedule and hardware.

\section{Experimental Setup}\label{sec:setup}
\subsection{Datasets and Evaluation Protocol}
\label{ssec:datasets}

We evaluate on the public infrared SR benchmark introduced with PSRGAN~\cite{huang2021psrgan} and adopted by IRSRMamba~\cite{huang2025irsrmamba} and GPSMamba~\cite{huang2025gpsmamba}: two test sets of $22$ images each, \emph{results-A} and \emph{results-C}, produced by the infrared--visible fusion methods of Liu \emph{et al.}~\cite{liu2018fusioncnn} and Zhang \emph{et al.}~\cite{zhang2017fusion}. The test images are the archives \texttt{results-A.zip} and \texttt{results-C.zip} of the Figshare dataset released with IRSRMamba~\cite{huang2024figshare_irsrmamba}; the same archives, together with \texttt{IR100.zip}, form the original PSRGAN release~\cite{huang2021figshare_psrgan}. We use these files without modification. The benchmark's ten-image CVC10 subset cannot be reported for any newly evaluated method because its image selection was never released; Sec.~\ref{ssec:panels} introduces reproducible test sets instead.

The training protocol~\cite{huang2025irsrmamba,huang2025gpsmamba} uses the first $265$ infrared images, in filename order, of the M3FD fusion subset~\cite{liu2022tardal}. Each high-resolution (HR) image is first cropped so that its height and width are multiples of the scale factor (the \texttt{modcrop} step of BasicSR), and the low-resolution (LR) input is then produced by MATLAB-compatible antialiased bicubic downsampling; our LR files are byte-identical to the officially released ones. Metrics follow the BasicSR convention of both prior works: PSNR, SSIM~\cite{wang2004ssim} and MSE on the Y channel of the BT.601 YCbCr transform, with a border equal to the scale factor removed. The metric also carries a fixed offset that every method shares: the test images have a single band, and the RGB-to-Y conversion rescales intensities by $219/255$ while the peak value stays at $255$, so reported PSNR is exactly $20\log_{10}(255/219)=1.32$\,dB above plain grayscale PSNR. The reference models replicate the grayscale input to three channels whereas TIRMamba uses the single channel directly; under the Y-channel metric the two settings carry the same information.

\subsection{Additional Test Sets}
\label{ssec:panels}

Both official test sets are fusion products, so we add three thermal test sets of increasing independence from the training data, all degraded with the identical bicubic pipeline and scored with the same Y-channel protocol. \textbf{M3FD-heldout} contains the $35$ raw thermal frames that follow the $265$ training images in filename order (images $266$--$300$); no method used them for training or model selection, they share the sensor and campaign of the training images, and they served as the development set for the TIRMamba-Rad diagnosis of Sec.~\ref{ssec:secondgen}. \textbf{HIT-UAV}~\cite{suo2023hituav} contributes $40$ frames drawn uniformly from its official test split ($640\times512$ pixels); no compared model has seen aerial imagery. \textbf{LLVIP}~\cite{jia2021llvip} contributes $40$ frames from its official test split, recorded by a fixed-mount street-surveillance long-wave-infrared camera that shares nothing with the training corpora. A fourth test set of $40$ optical aerial images from AID~\cite{xia2017aid} is summarized at the end of Sec.~\ref{ssec:optical}.

\subsection{Measurement Protocol}
\label{ssec:floor}

Benchmarks of this size have a noise floor below which differences cannot be interpreted. From the validation histories of our own runs, results-A PSNR spans $0.108$\,dB at $\times2$ and $0.011$\,dB at $\times4$ across the ten checkpoints around the selected one, whereas two runs that differ only in random seed end $0.010$\,dB apart, so most of the variation comes from checkpoint selection. We adopt this within-run spread as the resolution of the benchmark and treat any margin below it as a tie. Wherever per-image outputs exist we add two-sided paired Wilcoxon signed-rank tests; the family of $52$ tests over $26$ comparisons uses a Bonferroni-corrected threshold~\cite{dunn1961multiple,bland1995bonferroni} of $p<9.6\times10^{-4}$, and every test, favorable or not, is listed in Table~S-I of the supplementary material. Following the official codebases, checkpoints are selected on the test sets, which biases our numbers and the published ones in the same way; moving the selection of both final models to M3FD-heldout changes PSNR by at most $0.015$\,dB.

\subsection{Implementation Details}
\label{ssec:implementation}

\textbf{Pre-training and fine-tuning.} The network is pre-trained on $68{,}915$ grayscale sub-images cut from the $800$ DIV2K~\cite{agustsson2017div2k} training images, with inputs normalized as $(x-0.5)/0.5$; this checkpoint is the zero-shot row of Table~\ref{tab:ablation_training}. The main model is then fine-tuned from it on the mixture of Sec.~\ref{ssec:training} with LR patches of $64$ pixels (HR $128$), batch size $8$, Adam with $\beta=(0.9,0.99)$, a learning rate of $2\times10^{-5}$ halved at $15$k and $25$k iterations, $30$k iterations in total, the loss of (\ref{eq:loss}), an exponential moving average of the weights (decay $0.995$) used for evaluation, and seed $3407$ as in the benchmark protocol; the best checkpoint is at $8$k iterations. The \emph{transfer} variant uses the identical schedule on M3FD-265 alone, following the practice of PSRGAN~\cite{huang2021psrgan} (best at $10$k). All training and timing use one NVIDIA GeForce RTX~3080 GPU (10\,GB).

\textbf{$\times4$ models.} The $\times4$ replay model warm-starts from the best $\times2$ replay checkpoint; its new reconstruction head is trained at $2\times10^{-4}$ and the backbone at $2\times10^{-5}$, with LR patches of $32$, batch size $16$, milestones at $20$k and $32$k, and $40$k iterations (best at $13$k). The replay-free $\times4$ transfer variant warm-starts from the LR-patch-$16$, $L_1$ $\times2$ transfer checkpoint and fine-tunes on M3FD-265 alone under $L_1$; the visible-only $\times4$ model uses the $\times4$ schedule above but draws every patch from DIV2K.

\textbf{TIRMamba-Rad and the intermediate configuration.} TIRMamba-Rad is trained from scratch on the same mixed stream at LR patch $64$ for $350$k iterations (best checkpoint at $330$k), and its $\times4$ model warm-starts from that checkpoint under the $\times4$ schedule; the two final models therefore differ in schedule as well as in architecture. The $912.9$K intermediate configuration that serves as the ablation base is trained from scratch for $500$k iterations with the patch size raised from $32$ to $64$ and read at $260$k.

\textbf{Training data.} Our final models see grayscale DIV2K plus M3FD-265; the strict-protocol control sees M3FD-265 alone. The compared methods were not retrained with external data, so Table~\ref{tab:x2} compares our disclosed training procedure against their reported protocol numbers, with Table~\ref{tab:ablation_training} supplying the matched-data evidence; GPSMamba also uses external visual knowledge through the ImageNet-pre-trained VGG19 in its loss~\cite{huang2025gpsmamba}.

\subsection{Reproduced Baselines}
\label{sec:local_repro}
The competitors' reported numbers cover only the fusion test sets and the fidelity metrics, so for the additional test sets, the perceptual metrics and the qualitative figures we run the officially released checkpoints of IRSRMamba, GPSMamba and PSRGAN locally under each official test configuration (for the perceptual metrics on the fusion sets, IRSRMamba is scored on the super-resolved images its authors released). Rows derived from these outputs are marked $^{*}$; the reproduced IRSRMamba and GPSMamba outputs match the reported PSNR to within $0.030$ and $0.005$\,dB, respectively. Table~\ref{tab:efficiency} and the parameter ratios of Sec.~\ref{sec:generalization} use the parameter counts instantiated by the released checkpoints ($20{,}422$K\,/\,$20{,}570$K for IRSRMamba and $36{,}795$K\,/\,$36{,}943$K for GPSMamba at $\times2$\,/\,$\times4$), whereas Table~\ref{tab:x2} quotes the reported counts. Only the original $313$K\,/\,$350$K PSRGAN weights are public, so rows computed from them carry the $^{\ddagger}$ marker; the supplementary material gives the reproduction details.

\section{Results on the Official Benchmark}\label{sec:experiments}

\begin{table*}[!t]
\centering
\caption{Quantitative comparison on the official benchmark at $\times2$ and $\times4$ (PSNR$\uparrow$\,/\,SSIM$\uparrow$, Y channel, crop border equal to the scale factor). Competitor rows are the reported protocol numbers at the configuration and parameter count reported in the source benchmark (eleven of the twenty reported configurations; the full set with MSE, FLOPs and latency is in Tables~S-VII and~S-VIII of the supplementary material); the anomalous MambaIRv2 $\times4$ entry is carried over unchanged. Markers, used in all tables: $^{\dagger}$\,our replay-mixed training; $^{\P}$\,strict-protocol control trained on M3FD-265 alone. \textbf{Bold}: best among protocol-trained methods; \underline{underline}: second. Margins inside the noise floor of Sec.~\ref{ssec:floor} are ties (Sec.~\ref{ssec:sota}).}
\label{tab:x2}\label{tab:x4}
\setlength{\tabcolsep}{3.4pt}
\footnotesize
\begin{tabular}{l r c c c c c c c c}
\toprule
\multirow{3}{*}{Method} & \multirow{3}{*}{\#Params (K)} & \multicolumn{4}{c}{$\times2$} & \multicolumn{4}{c}{$\times4$}\\
\cmidrule(lr){3-6}\cmidrule(lr){7-10}
 & & \multicolumn{2}{c}{results-A} & \multicolumn{2}{c}{results-C} & \multicolumn{2}{c}{results-A} & \multicolumn{2}{c}{results-C}\\
 & & PSNR & SSIM & PSNR & SSIM & PSNR & SSIM & PSNR & SSIM\\
\midrule
Bicubic                                   &      -- & 37.6846 & 0.9270 & 38.6222 & 0.9391 & 33.2435 & 0.8313 & 33.8318 & 0.8504\\
PSRGAN~\cite{huang2021psrgan}             &  2{,}414 & 39.2146 & 0.9429 & 40.0543 & 0.9539 & 34.4595 & 0.8540 & 35.1023 & 0.8715\\
SwinIR~\cite{liang2021swinir}             & 11{,}752 & 38.6899 & 0.9374 & 39.5215 & 0.9492 & 34.4321 & 0.8537 & 35.0329 & 0.8710\\
ShuffleMixer-T~\cite{sun2023shufflemixer} &     108 & 39.0465 & 0.9414 & 39.8766 & 0.9527 & 34.5440 & 0.8550 & 35.1640 & 0.8723\\
ATD-light~\cite{zhang2024atd}                   &     753 & 39.0453 & 0.9432 & 40.0375 & 0.9542 & 34.6113 & 0.8569 & 35.2347 & 0.8737\\
CATANet~\cite{liu2025catanet}             &     477 & 39.0886 & 0.9430 & 40.0064 & 0.9539 & 34.0613 & 0.8447 & 34.8947 & 0.8663\\
MambaIR~\cite{guo2024mambair}             & 20{,}421 & 39.1761 & 0.9437 & 40.1399 & 0.9544 & 34.0267 & 0.8510 & 34.5662 & 0.8681\\
MambaIRv2~\cite{guo2025mambairv2}         & 22{,}903 & 39.0421 & 0.9423 & 39.9295 & 0.9531 & 29.5295 & 0.8420 & 28.6532 & 0.8577\\
VisionMamba~\cite{zhu2024vim}             & 27{,}880 & 38.7805 & 0.9392 & 39.6339 & 0.9506 & 34.5941 & 0.8564 & 35.2327 & 0.8733\\
MambaOut~\cite{yu2025mambaout}            &  9{,}669 & 38.6375 & 0.9371 & 39.4900 & 0.9493 & 34.4483 & 0.8527 & 35.0456 & 0.8698\\
IRSRMamba~\cite{huang2025irsrmamba}       & 26{,}462 & \underline{39.3489} & \underline{0.9440} & 40.2302 & \underline{0.9548} & 34.6755 & 0.8577 & 35.3074 & \underline{0.8745}\\
GPSMamba~\cite{huang2025gpsmamba}         & 36{,}942 & \textbf{39.3505} & \underline{0.9440} & \underline{40.2418} & \underline{0.9548} & \underline{34.7421} & \underline{0.8587} & \underline{35.4007} & \textbf{0.8756}\\
\midrule
TIRMamba (strict protocol)$^{\P}$ (ours)  & 896\,/\,910 & 39.1937 & 0.9431 & 40.0711 & 0.9540 & 34.4077 & 0.8533 & 35.0550 & 0.8708\\
TIRMamba$^{\dagger}$ (ours)               & 896\,/\,910 & 39.2971 & \textbf{0.9443} & \textbf{40.2909} & \textbf{0.9549} & \textbf{34.7478} & \textbf{0.8588} & \textbf{35.4057} & \textbf{0.8756}\\
\bottomrule
\end{tabular}
\end{table*}

\subsection{Two Properties of the Protocol That Shape the Training Strategy}
\label{ssec:protocol_effects}\label{sec:transfer}\label{ssec:zeroshot}\label{ssec:domain_limited}\label{ssec:matched}\label{ssec:forgetting}\label{ssec:tile}\label{ssec:implications}

Eight third-party lightweight architectures of $228$K to $878$K parameters, released as RGB models trained on natural images, were tested under the protocol without any fine-tuning (Table~S-II of the supplementary material). On results-A at $\times2$ they span just $0.041$\,dB and come within $0.08$ to $0.12$\,dB of the $36.9$M GPSMamba trained under the protocol; at $\times4$, ATD-light reaches $34.7493$\,dB without ever seeing a thermal image, above both protocol-trained Mamba models. Our DIV2K-only TIRMamba behaves the same way, reaching $40.2767$\,dB on results-C before any infrared training, above GPSMamba's reported score (zero-shot row of Table~\ref{tab:ablation_training}), whereas the same architecture trained from scratch on the $265$ protocol images for $500$k iterations (Table~\ref{tab:ablation_training}) reaches only $39.0587$ and $39.9128$\,dB. The two test sets are fusion products~\cite{liu2018fusioncnn,zhang2017fusion} whose local statistics resemble those of natural grayscale images, so natural-image training transfers with almost no loss.

To separate what the $265$ images contribute as thermal data from what they contribute as additional data, we adapt the same pre-trained network twice under identical settings, on the $265$ protocol images and on $265$ natural images (middle block of Table~\ref{tab:ablation_training}). On results-A the infrared corpus gains $0.278$\,dB against $0.103$\,dB for the natural one, and on the raw-thermal frames of Sec.~\ref{ssec:heldout} the gap is larger ($51.0068$ against $50.7230$\,dB). Both corpora lower results-C by similar small amounts ($-0.047$ and $-0.018$\,dB), a small-corpus effect that the replay mixture is meant to address; two asymmetries of the control make these gaps upper bounds (supplementary material).

The second effect is isolated with tiled inference. The model fine-tuned at LR patch $16$ scores $39.2115$\,dB on whole images but $39.3177$\,dB on $48$-pixel tiles, a gain of $0.106$\,dB that comes only from matching the test-time input statistics to those of training. Each scan branch visits every pixel, so the sequence length grows from $256$ tokens for a $16\times16$ patch to $19.6$K--$110.6$K tokens on the official $\times2$ test images ($140\times140$ to $384\times288$ pixels; $4.9$K--$27.6$K at $\times4$), which shifts the state statistics that control the selective decay horizon~\cite{gu2023mamba}. Under orthonormal normalization the input magnitude of the spectral branch also grows approximately as $\sqrt{HW}$. Retraining at LR patch $64$ reduces the largest tiling gain to $0.019$\,dB (full grid in the supplementary material); tiling is permitted by the protocol and used in the released IRSRMamba test code~\cite{huang2025irsrmamba}, whereas all TIRMamba scores use whole-image inference. A global operator should therefore be trained with matched statistics, as our training strategy does, or made size-invariant by design; the intermediate configuration takes the second route, with the consequences analyzed in Sec.~\ref{ssec:secondgen}.

\subsection{Comparison with the State of the Art}
\label{ssec:sota}\label{ssec:perceptual}

Table~\ref{tab:x2} compares TIRMamba with the methods of the source benchmark, nineteen prior methods in twenty configurations (eleven shown; the full set is in the supplementary material). Every competitor row keeps the configuration and parameter count reported in the source benchmark, where each was retrained under the same M3FD-265 protocol; the MambaIRv2 $\times4$ entry, whose PSNR falls below bicubic while its SSIM stays competitive, suggests an unconverged run there and is carried over unchanged. For brevity we call the two state-space methods at the top of the published results, IRSRMamba and GPSMamba, the \emph{large reference models}.

\textbf{Scale factor four.} TIRMamba$^{\dagger}$ obtains the best value in all four $\times4$ columns with $910$K parameters. We regard this as parity with the best published model rather than a lead: the margins over GPSMamba ($0.006$ and $0.005$\,dB) lie inside the $0.011$\,dB noise floor of Sec.~\ref{ssec:floor} and are not significant per image ($p=0.46$/$0.26$), whereas the margins over IRSRMamba ($0.072$ and $0.098$\,dB) exceed the floor. TIRMamba reaches these scores with $29$ to $40$ times fewer parameters (Fig.~\ref{fig:scatter}).  Even without replay, the transfer variant reaches $34.6245$ and $35.2701$\,dB, within $0.051$\,dB of IRSRMamba.

\textbf{Scale factor two.} TIRMamba obtains the best SSIM on both test sets and the best PSNR on results-C. On results-A it is below only the two reference models, and its $0.052$\,dB gap to IRSRMamba is half the checkpoint-selection spread and indistinguishable from zero on a per-image basis ($p=0.80$). The transfer variant of Table~\ref{tab:ablation_training} brings results-C to within $0.0004$\,dB of IRSRMamba's reported score, so the two models match under the transfer setting of PSRGAN, with $30$ times fewer parameters. 

\textbf{Architecture alone.} The strict-protocol row trains TIRMamba on the $265$ protocol images alone and reaches $39.1937$/$40.0711$\,dB at $\times2$, above $17$ of the $20$ reported entries on each test set (Table~S-VII) with $896$K parameters, and $34.4077$/$35.0550$\,dB at $\times4$; the training strategy of Sec.~\ref{ssec:training} accounts for the remaining gain.

\textbf{Per-image tests.} At $\times2$, the results-C PSNR advantages over both reproduced reference models are significant and frame-consistent ($+0.051$ and $+0.086$\,dB, $19$ of $22$ frames each, $p=1.8\times10^{-4}$), as are the SSIM advantages on results-A ($p=1.2\times10^{-5}$ against both), whereas the results-C SSIM margins ($p=0.004$/$0.013$) do not survive the family correction; the results-A PSNR differences are not significant ($p=0.78$/$0.80$), and at $\times4$ no difference against the reproduced GPSMamba is significant ($p=0.46$/$0.26$). Of the $52$ tests in Table~S-I, $31$ survive the corrected threshold, $25$ in our favor and $6$ against. The six come from ATD-light fine-tuned with our own training strategy, which reaches $39.3169$/$40.3184$\,dB at $\times2$ and $34.7871$/$35.4551$\,dB at $\times4$, above TIRMamba$^{\dagger}$ on all four fusion-set entries, and from the transfer variant on HIT-UAV at $\times2$ (Sec.~\ref{ssec:hituav}). On these test sets the training strategy changes the score more than the backbone does, which is why our claim on the fusion benchmark is parity at a sub-million parameter count rather than a lead.

\textbf{Perceptual quality.} Table~S-III of the supplementary material reports LPIPS, DISTS, NIQE, BRISQUE, CLIP-IQA and DBCNN for every model: at $\times2$ TIRMamba gives lower LPIPS, NIQE and BRISQUE than both reference models on both test sets, whereas at $\times4$ the reference models lead those metrics and TIRMamba leads the two learned no-reference metrics; the adversarially trained PSRGAN trades $1.5$--$1.9$\,dB of PSNR for the best LPIPS, DISTS and BRISQUE, while the two learned no-reference metrics favor the fidelity-trained models~\cite{blau2018pd}.

\begin{table}[!t]
\centering
\caption{Efficiency measured by us on one RTX~3080 GPU under a single measurement setup (Sec.~\ref{ssec:efficiency}): instantiated parameters, thop FLOPs, full-image latency and peak GPU memory. $^{*}$\,validated local reproduction; $^{\ddagger}$\,released original PSRGAN weights. PSRGAN at its retrained $2{,}414$K size, VisionMamba and MambaOut are absent because no public configuration reproduces their reported parameter counts.}
\label{tab:efficiency}
\setlength{\tabcolsep}{3pt}
\footnotesize
\resizebox{\columnwidth}{!}{%
\begin{tabular}{l r c c c c c c}
\toprule
\multirow{2}{*}{Method} & \multirow{2}{*}{\#Params (K) $\times2$\,/\,$\times4$} & \multicolumn{3}{c}{$\times2$ ($135\times180$)} & \multicolumn{3}{c}{$\times4$ ($67\times90$)}\\
\cmidrule(lr){3-5}\cmidrule(lr){6-8}
 & & FLOPs (G) & Time (ms) & Mem.\ (MB) & FLOPs (G) & Time (ms) & Mem.\ (MB)\\
\midrule
ShuffleMixer-T~\cite{sun2023shufflemixer}    & 108\,/\,113           & 2.6   & 8.2    & 28    & 0.8   & 7.9   & 26\\
PSRGAN$^{\ddagger}$~\cite{huang2021psrgan}   & 313\,/\,350           & 12.9  & 6.1    & 93    & 9.6   & 4.6   & 79\\
CATANet~\cite{liu2025catanet}                & 477\,/\,535           & 11.5  & 63.5   & 843   & 4.3   & 31.0  & 250\\
ATD-light~\cite{zhang2024atd}                      & 753\,/\,769           & 19.3  & 132.2  & 349   & 5.5   & 63.9  & 108\\
MambaIR~\cite{guo2024mambair}                & 20{,}422\,/\,20{,}570 & 401.7 & 354.1  & 955   & 103.4 & 92.0  & 353\\
MambaIRv2~\cite{guo2025mambairv2}            & 22{,}584\,/\,22{,}732 & 588.7 & 604.8  & 845   & 168.2 & 179.7 & 368\\
IRSRMamba$^{*}$~\cite{huang2025irsrmamba}    & 20{,}422\,/\,20{,}570 & 401.7 & 354.0  & 955   & 103.4 & 92.1  & 353\\
GPSMamba$^{*}$~\cite{huang2025gpsmamba}      & 36{,}795\,/\,36{,}943 & 857.3 & 1594.6 & 3664  & 243.1 & 312.4 & 635\\
\midrule
TIRMamba$^{\dagger}$ (ours) & 896\,/\,910 & 16.3 & 35.9 & 136 & 4.1 & 33.1 & 45\\
TIRMamba-Rad$^{\dagger}$ (ours) & 948\,/\,990 & 20.9 & 72.0 & 172 & 5.5 & 49.1 & 55\\
\bottomrule
\end{tabular}}
\end{table}

\subsection{Efficiency and Qualitative Results}
\label{ssec:efficiency}\label{ssec:qualitative}

Table~\ref{tab:efficiency} compares parameters, FLOPs, full-image latency and peak GPU memory for eight of the benchmark architectures and both of our models (the remaining measured rows are in Tables~S-VII and~S-VIII of the supplementary material), all measured by us on one RTX~3080 GPU under a single measurement setup; neither Mamba paper reports FLOPs or latency. Rows without released weights are built from the official code at the benchmark configuration and kept only when the built parameter count matches the reported one (details in the supplementary material). PSRGAN at its retrained $2{,}414$K size, VisionMamba and MambaOut have no public configuration that reproduces the reported count, so they have no row; measuring a guessed configuration would mean measuring a model other than the one whose accuracy is quoted. At $\times2$, TIRMamba processes a $135\times180$ input in $35.9$\,ms, $9.9\times$ faster than IRSRMamba and $44\times$ faster than GPSMamba, with $25\times$/$53\times$ fewer FLOPs and $7\times$/$27\times$ lower peak memory; the $\times4$ gaps are $2.8\times$/$9.4\times$. TIRMamba-Rad costs about twice the base latency at one-twentieth of IRSRMamba's FLOPs.

\begin{figure*}[!t]
\centering
\includegraphics[width=0.98\textwidth]{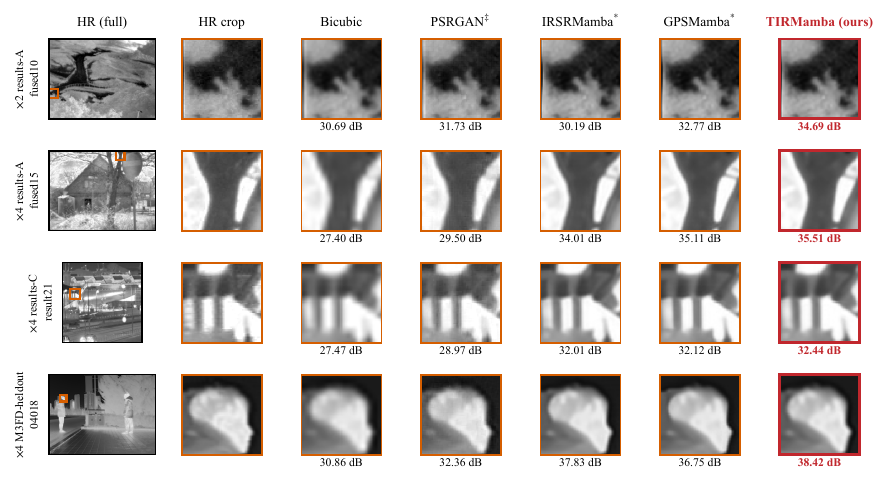}
\caption{Visual comparison at $\times2$ (top row) and $\times4$ (remaining rows); the bottom row is a raw thermal M3FD-heldout frame. Each row shows the full HR frame with the crop location, the HR crop, and the method crops annotated with per-crop PSNR (crop rule in Sec.~\ref{ssec:qualitative}).}
\label{fig:visual}
\end{figure*}

\textbf{Qualitative results.} Fig.~\ref{fig:visual} shows results at both scales against bicubic and the reproduced competitors, with per-crop PSNR. The crop is placed by an automatic rule that maximizes HR texture variance weighted by bicubic error, so it never consults any method's output. The frames shown are ones on which our model leads on that crop; they represent the $9$ to $14$ frames per test set on which that holds, not the averages. TIRMamba renders edge transitions more sharply than bicubic and at least as well as the much larger reference models.

\section{Generalization Beyond the Fusion Benchmark}
\label{sec:generalization}

\begin{table*}[!t]
\centering
\caption{Generalization to the three thermal test sets of Sec.~\ref{ssec:panels}: raw-thermal M3FD-heldout, zero-shot aerial HIT-UAV~\cite{suo2023hituav}, and the independent surveillance sensor LLVIP~\cite{jia2021llvip} (PSNR$\uparrow$\,/\,SSIM$\uparrow$; markers as in Tables~\ref{tab:x2} and~\ref{tab:efficiency}; parameters and latency in Table~\ref{tab:efficiency}). $^{\S\S}$\,trained on a different raw-M3FD split with possible frame overlap, hence unranked (Sec.~\ref{ssec:heldout}).}
\label{tab:heldout}\label{tab:llvip}
\setlength{\tabcolsep}{3.6pt}
\footnotesize
\begin{tabular}{l c c c c c c c c c c c c}
\toprule
\multirow{3}{*}{Method} & \multicolumn{4}{c}{M3FD-heldout (raw thermal)} & \multicolumn{4}{c}{HIT-UAV (zero-shot aerial)} & \multicolumn{4}{c}{LLVIP (independent sensor)}\\
\cmidrule(lr){2-5}\cmidrule(lr){6-9}\cmidrule(lr){10-13}
 & \multicolumn{2}{c}{$\times2$} & \multicolumn{2}{c}{$\times4$} & \multicolumn{2}{c}{$\times2$} & \multicolumn{2}{c}{$\times4$} & \multicolumn{2}{c}{$\times2$} & \multicolumn{2}{c}{$\times4$}\\
 & PSNR & SSIM & PSNR & SSIM & PSNR & SSIM & PSNR & SSIM & PSNR & SSIM & PSNR & SSIM\\
\midrule
Bicubic & 46.2755 & 0.9919 & 36.7248 & 0.9450 & 44.3950 & 0.9868 & 36.4883 & 0.9381 & 45.3021 & 0.9935 & 36.2073 & 0.9651\\
PSRGAN$^{\ddagger}$~\cite{huang2021psrgan} & 45.8364 & 0.9861 & 36.5073 & 0.9370 & 43.6266 & 0.9786 & 36.5489 & 0.9265 & 44.3612 & 0.9860 & 34.5795 & 0.9456\\
IRSRMamba$^{*}$~\cite{huang2025irsrmamba} & \underline{50.9925} & \underline{0.9960} & \underline{41.2956} & \underline{0.9718} & \underline{47.0329} & \textbf{0.9915} & 39.5859 & 0.9591 & 49.5494 & 0.9957 & 40.9524 & 0.9822\\
GPSMamba$^{*}$~\cite{huang2025gpsmamba} & 50.9514 & \underline{0.9960} & 41.0734 & 0.9711 & \textbf{47.0719} & \textbf{0.9915} & \underline{39.6305} & \underline{0.9598} & \underline{49.7891} & \underline{0.9958} & 41.0220 & 0.9831\\
CoRPLE-light$^{\S\S}$~\cite{huang2024corple} & 51.3143 & 0.9963 & 41.8309 & 0.9744 & 46.1999 & 0.9906 & 39.5037 & 0.9591 & -- & -- & -- & --\\
\midrule
TIRMamba$^{\dagger}$ (ours) & 50.8504 & \underline{0.9960} & 41.0540 & 0.9706 & 46.9282 & \underline{0.9910} & 39.5686 & 0.9589 & 49.3758 & \textbf{0.9959} & \underline{41.1459} & \underline{0.9834}\\
TIRMamba-Rad$^{\dagger}$ (ours) & \textbf{51.3310} & \textbf{0.9962} & \textbf{41.3850} & \textbf{0.9723} & 46.9564 & \textbf{0.9915} & \textbf{39.7452} & \textbf{0.9606} & \textbf{49.8351} & \textbf{0.9959} & \textbf{41.1687} & \textbf{0.9837}\\
\bottomrule
\end{tabular}
\end{table*}

Table~\ref{tab:heldout} evaluates every model on the three thermal test sets of Sec.~\ref{ssec:panels}, which remove the fusion step, change the platform, and change the sensor, in that order; margins are read against the per-image tests of Table~S-I where available.

\subsection{Raw Thermal Imagery (M3FD-heldout)}
\label{ssec:heldout}

The raw-thermal set separates methods far more clearly than the fusion sets: the gap between bicubic and the best method is $5.1$\,dB at $\times2$, against $1.7$\,dB on results-A. At $\times2$, TIRMamba-Rad obtains the best PSNR among protocol-trained methods, $51.3310$\,dB (Sec.~\ref{ssec:secondgen}). The replay-free transfer variant of Table~\ref{tab:ablation_training} reaches $50.9996$\,dB, $0.007$\,dB above the reproduced IRSRMamba, within our reproduction tolerance and not significant ($p=0.13$), so the two are at PSNR parity with $1/23$ of the parameters; the replay model loses $0.15$\,dB here, mirroring its gains on the fusion sets. At $\times4$, TIRMamba$^{\dagger}$ is $0.019$\,dB behind GPSMamba ($p=0.47$) and $0.24$\,dB behind IRSRMamba, and TIRMamba-Rad gives the highest score among the ranked models. PSRGAN$^{\ddagger}$ scores below bicubic at both scales, so an adversarial visible-band prior does not transfer to raw thermal data. CoRPLE-light was trained on a raw-M3FD $182$-image split that may overlap these frames and is therefore listed without ranking. The bottom row of Fig.~\ref{fig:visual} is a $\times4$ example from this set.

\subsection{UAV Thermal Imagery (HIT-UAV)}
\label{ssec:hituav}

The platform shift is more discriminative than the fusion sets. At $\times2$ every ranked fidelity-oriented model lies within $0.15$\,dB: GPSMamba is first at $47.0719$\,dB and TIRMamba-Rad$^{\dagger}$ reaches $46.9564$\,dB, $0.12$\,dB lower at $1/39$ of the parameters and $22\times$ the speed (Table~\ref{tab:efficiency}). At $\times4$ the ordering changes. TIRMamba-Rad$^{\dagger}$ obtains the highest score, $39.7452$\,dB, with margins of $+0.11$\,dB over GPSMamba on $36$ of $40$ frames and $+0.16$\,dB over IRSRMamba on all $40$ frames (both $p<10^{-4}$), and it is higher on all $40$ frames in SSIM ($p=1.8\times10^{-12}$). At $\times2$ the transfer variant trails GPSMamba on $37$ of $40$ frames ($-0.090$\,dB, $p=1.2\times10^{-9}$), the only cross-sensor gap that survives the correction.

\subsection{An Independent Thermal Sensor (LLVIP)}
\label{ssec:llvip}\label{ssec:optical}

This test set shares nothing with any training corpus, and the fusion-benchmark ranking does not carry over. At $\times4$, both of our models score above both reference models. The $910$K base model gains $0.19$\,dB over IRSRMamba ($32$ of $40$ frames, $p=1.6\times10^{-4}$) and $0.12$\,dB over GPSMamba, a PSNR margin that is significant before but not after the family correction while its SSIM margin survives on $33$ of $40$ frames, and TIRMamba-Rad exceeds GPSMamba by $0.147$\,dB on $28$ of $40$ frames ($p=2.0\times10^{-5}$). At $\times2$, TIRMamba-Rad and GPSMamba differ by $0.046$\,dB, a tie after correction ($p=0.024$) at $1/39$ of the parameters and $1/22$ of the latency, with an SSIM advantage on all $40$ frames ($p=1.8\times10^{-12}$); against IRSRMamba, TIRMamba-Rad leads by $0.286$\,dB on $39$ of $40$ frames ($p<10^{-9}$). The $\times2$ base model is the only entry that generalizes less well, $0.174$\,dB behind IRSRMamba ($p=0.10$) with an SSIM advantage on $38$ of $40$ frames; this gap is consistent with the raw-thermal diagnosis of Sec.~\ref{ssec:secondgen}, which was made on M3FD-heldout alone, and closes at $\times4$.

\textbf{Optical imagery after adaptation.} On $40$ grayscale optical aerial images from AID~\cite{xia2017aid}, degraded and scored as the thermal test sets, the final $\times2$ and $\times4$ models stay within $0.026$\,dB of the visible-only models they were fine-tuned from, whereas fine-tuning on M3FD-265 alone costs $0.360$\,dB (Table~S-V of the supplementary material), so the replay strategy preserves the visible-band accuracy of the pre-trained network.

\section{Ablation Studies}\label{sec:analysis}

\begin{table}[!t]
\centering
\caption{Training-strategy ablation at $\times2$ (PSNR$\uparrow$/SSIM$\uparrow$). All rows share the architecture and differ in adaptation data, patch size and loss (Sec.~\ref{ssec:training_ablation}); the from-scratch row uses $[0,1]$ input normalization rather than the $(x{-}0.5)/0.5$ of the pre-trained models.}
\label{tab:ablation_training}
\setlength{\tabcolsep}{2.2pt}
\footnotesize
\resizebox{\columnwidth}{!}{%
\begin{tabular}{l c c}
\toprule
Training strategy & results-A & results-C\\
\midrule
Zero-shot (DIV2K only, no adaptation) & 39.0305/0.9441 & 40.2767/\textbf{0.9549}\\
From scratch (M3FD only, 500k)  & 39.0587/0.9420 & 39.9128/0.9530\\
Fine-tune on M3FD only, patch 16             & 39.2115/0.9431 & 40.0940/0.9540\\
Fine-tune on M3FD only, patch 64 (transfer)  & \textbf{39.3078}/0.9440 & 40.2298/0.9548\\
\midrule
\multicolumn{3}{l}{\emph{Matched-corpus control (Sec.~\ref{ssec:protocol_effects})}}\\
\quad 265 natural images        & 39.1337/0.9442 & 40.2585/\textbf{0.9549}\\
\quad 265 infrared images       & 39.3086/0.9440 & 40.2294/0.9548\\
\midrule
Replay-mixed, patch 64 (ours) & 39.2971/\textbf{0.9443} & \textbf{40.2909}/\textbf{0.9549}\\
\bottomrule
\end{tabular}}
\end{table}

\subsection{Training Strategy}
\label{ssec:training_ablation}

Table~\ref{tab:ablation_training} varies the adaptation data, the LR patch size and the loss on a single architecture, with the matched-corpus runs of Sec.~\ref{ssec:protocol_effects} as its middle block. Increasing the LR patch size from $16$ to $64$ on pure M3FD raises results-A to $39.3078$\,dB, the best M3FD-only value in Table~\ref{tab:ablation_training}, while results-C stays at $40.2298$\,dB, below the zero-shot starting point. Experience replay~\cite{rolnick2019experience} then removes the trade-off, giving $39.2971$ and $40.2909$\,dB, a gain of $0.061$\,dB on results-C at a cost of $0.011$\,dB on results-A, together with the highest SSIM in Table~\ref{tab:x2}. The patch-$16$ rows also use $L_1$ instead of (\ref{eq:loss}), and the matched-control infrared run repeats the patch-$64$ infrared configuration at the same seed, agreeing with it to $0.0008$/$0.0004$\,dB.

\begin{table}[!t]
\centering
\caption{Component ablation at $\times2$ (PSNR$\uparrow$/SSIM$\uparrow$) on the $912.9$K intermediate configuration (Sec.~\ref{ssec:plus}; Table~S-IV). All rows share one $60$k mixed-domain from-scratch protocol and are evaluated at the same window, the mean over the validations at $50$k, $55$k and $60$k; $\sigma$ is each row's own PSNR standard deviation over that window, and a margin counts as an effect only when it exceeds twice the two rows' combined $\sigma$.  \textbf{Bold}: best per column.}
\label{tab:ablation_module}
\setlength{\tabcolsep}{2.6pt}
\footnotesize
\resizebox{\columnwidth}{!}{%
\begin{tabular}{l r c c c}
\toprule
Variant & \#Par.\,(K) & results-A & results-C & $\sigma_{A}$/$\sigma_{C}$\\
\midrule
Full model                  &   912.9 & 38.646/0.9436 & 39.388/0.9544 & 0.122/0.124\\
\midrule
raster scan (no serpentine) &   912.9 & 38.014/0.9423 & 37.824/0.9522 & 0.426/1.639\\
w/o detail branch (DDB)     &   894.5 & 35.574/0.9066 & 36.910/0.9371 & 0.201/0.478\\
untied bi-SSM               &  1132.1 & 37.277/0.9431 & 38.042/0.9538 & 0.497/0.551\\
$L_1$ loss (no FFT term)    &   912.9 & 38.820/0.9433 & 39.774/0.9542 & 0.057/0.026\\
\midrule
w/o Spectral Prior Branch   &   889.1 & 39.208/0.9436 & 40.174/0.9544 & 0.009/0.002\\
w/o Thermal Prior Highway   &   778.4 & \textbf{39.256}/\textbf{0.9437} & \textbf{40.178}/\textbf{0.9545} & 0.004/0.002\\
\bottomrule
\end{tabular}}
\end{table}

\subsection{Module Ablation}
\label{ssec:module_ablation}\label{ssec:ablbase}

All rows of Table~\ref{tab:ablation_module} share the $912.9$K intermediate configuration of Sec.~\ref{ssec:plus}, whose seven refinements over TIRMamba are listed in Table~S-IV of the supplementary material, with their parameter deltas in Sec.~S-IV; all are trained from scratch on the mixed stream under one $60$k protocol and evaluated at a common window.

Two components are necessary. Removing the detail branch costs $3.07$/$2.48$\,dB, an order of magnitude above the resolution threshold, which confirms that local convolution is the essential complement to a state-space scan~\cite{guo2024mambair}, and untying the bidirectional SSM weights adds $219$K parameters and still loses $1.37$/$1.35$\,dB. Raster scanning ($-0.63$/$-1.56$\,dB) shows the same trend but is among the least stable runs ($\sigma$ up to $1.64$\,dB), so its margin does not clear the bar. These conclusions carry over to the final models because the decisive effects are one to two orders of magnitude larger than the score differences of at most $0.05$\,dB between the configurations. Dropping the FFT term improves results-C by $0.386$\,dB under this protocol; the term was selected under the $30$k replay schedule and is kept on that basis.

The two variants that remove the TPH, or only the spectral prior, converge smoothly ($\sigma\le0.009$\,dB) and score higher on the fusion sets, by $0.61$/$0.79$ and $0.56$/$0.79$\,dB. Fusion-derived test sets reward natural-image statistics rather than the radiometric characteristics the priors target, so the priors can only be validated on raw thermal imagery, where the prior-conditioned mechanisms of TIRMamba-Rad do help (Secs.~\ref{ssec:heldout} and~\ref{ssec:secondgen}); the scale-invariant spectral prior is also an optimization liability under a short from-scratch schedule. Repeated on the stable spectral-prior-free configuration, the scan comparison resolves cleanly and goes against our design choice: raster order reaches $39.2683\pm0.0052$\,dB on results-A against $39.2077\pm0.0086$\,dB for stripe-serpentine, results-C is tied ($40.1722$ against $40.1743$), and the raw-thermal set also favors raster ($50.8989$ against $50.8069$\,dB), so stripe-serpentine order brings no accuracy benefit at this scale and is kept only for its locality argument, at equal cost.

\subsection{Raw-Thermal Failure Diagnosis and TIRMamba-Rad}
\label{ssec:secondgen}

\textbf{Diagnosis.} On M3FD-heldout at $\times2$, the intermediate-configuration checkpoint that is strongest on the fusion sets scores only $47.9885$\,dB, $3$\,dB below the base model's transfer variant ($50.9996$\,dB), while its SSIM ($0.9962$) is the highest among our own configurations, which indicates a sparse error of large amplitude. Test-time ablation localizes the cause to the scale-invariant spectral prior: biased convolutions in the forward-normalized branch respond even to near-zero-frequency content, which on the near-zero backgrounds of raw thermal frames produces constant impulses at the image corners and transients at the stripe boundaries. No test-time modification repairs it on both domains (zeroing the spectral template gains $0.48$\,dB on heldout but loses $0.84$\,dB on results-A), so the remedy is architectural (Sec.~\ref{ssec:plus}).

\textbf{Pilot study.} Table~S-IX of the supplementary material repeats the $60$k protocol one element at a time. The first-order hold improves all three sets at zero parameter cost and is adopted. Restoring $\beta$, the offset-injection route that the redesign removes, helps the fusion sets but costs $0.126$\,dB on raw thermal; removing PCS helps the fusion sets slightly while hurting raw thermal, so PCS is a mechanism with a documented raw-thermal benefit rather than a universal gain. An extension that also injects the prior into the input matrix $B$ was tested and not adopted (supplementary material). M3FD-heldout acts as a development set here, so checkpoint selection uses the fusion sets only.

\textbf{Full schedule.} TIRMamba-Rad ($948$K) reaches $39.3493$/$40.2548$\,dB on results-A/results-C, matching the published scores of both reference models ($39.3489$/$40.2302$ for IRSRMamba, $39.3505$/$40.2418$ for GPSMamba) to within the noise floor of Sec.~\ref{ssec:floor}. On M3FD-heldout it reaches $51.3310$\,dB, $0.34$\,dB above the reproduced IRSRMamba, and the $3$\,dB anomaly is gone at no cost on the fusion sets. The $\times4$ TIRMamba-Rad model ($989.8$K) gives the highest $\times4$ scores among the ranked models in Table~\ref{tab:heldout} while scoring $34.6865$ and $35.2842$\,dB on the $\times4$ fusion sets, $0.061$ and $0.122$\,dB below the base model. That small loss on fused imagery is the price of the raw-thermal specialization, which is why the base model is the primary model and TIRMamba-Rad a second operating point.

\subsection{Limitations}
\label{ssec:limitations}

Rankings on the two fusion test sets, ours included, should be read as ties unless the margin exceeds the noise floor of Sec.~\ref{ssec:floor}; checkpoint selection on the test sets follows the official codebases, and the resulting bias is bounded at $0.015$/$0.004$\,dB. M3FD-heldout shares its campaign with the training images and served as the development set for TIRMamba-Rad, so LLVIP provides the independent confirmation. Every fidelity test set uses the protocol's bicubic degradation; a no-reference probe on native-resolution HIT-UAV frames (supplementary material) does not separate the sub-million models from the large ones, and real-degradation fidelity needs field data that no public thermal-SR benchmark supplies. 

\section{Conclusion}\label{sec:conclusion}
This paper presented TIRMamba, a sub-million-parameter state-space network for infrared image super-resolution, together with a replay-mixed training strategy suited to the benchmark protocol. Thermal cues computed once at the input are carried through a Thermal Prior Highway that modulates a weight-tied bidirectional selective scan, and a dual-scale detail branch supplies the local structure a scan alone forgets. On the official benchmark the network matches the strongest published models at $\times4$ with $29$ to $40$ times fewer parameters, and across the two scales it runs $2.8$ to $44$ times faster. The ablations support the detail branch and the tied bidirectional weights, and show that the thermal priors help on raw thermal imagery rather than on the fusion test sets. A failure that appeared only on raw thermal data led to TIRMamba-Rad, which anchors the reconstruction to the radiometric level and conditions the selective-scan timescale on the thermal prior. It gives the best $\times4$ results on the raw-thermal, UAV and independent-sensor test sets; on the fusion benchmark it matches the reference models at $\times2$ and stays within $0.12$\,dB of the strongest published $\times4$ scores. 

\bibliographystyle{IEEEtran}
\bibliography{references}

\raggedbottom

\begin{IEEEbiography}[{\includegraphics[width=1in,height=1.25in,clip,keepaspectratio]{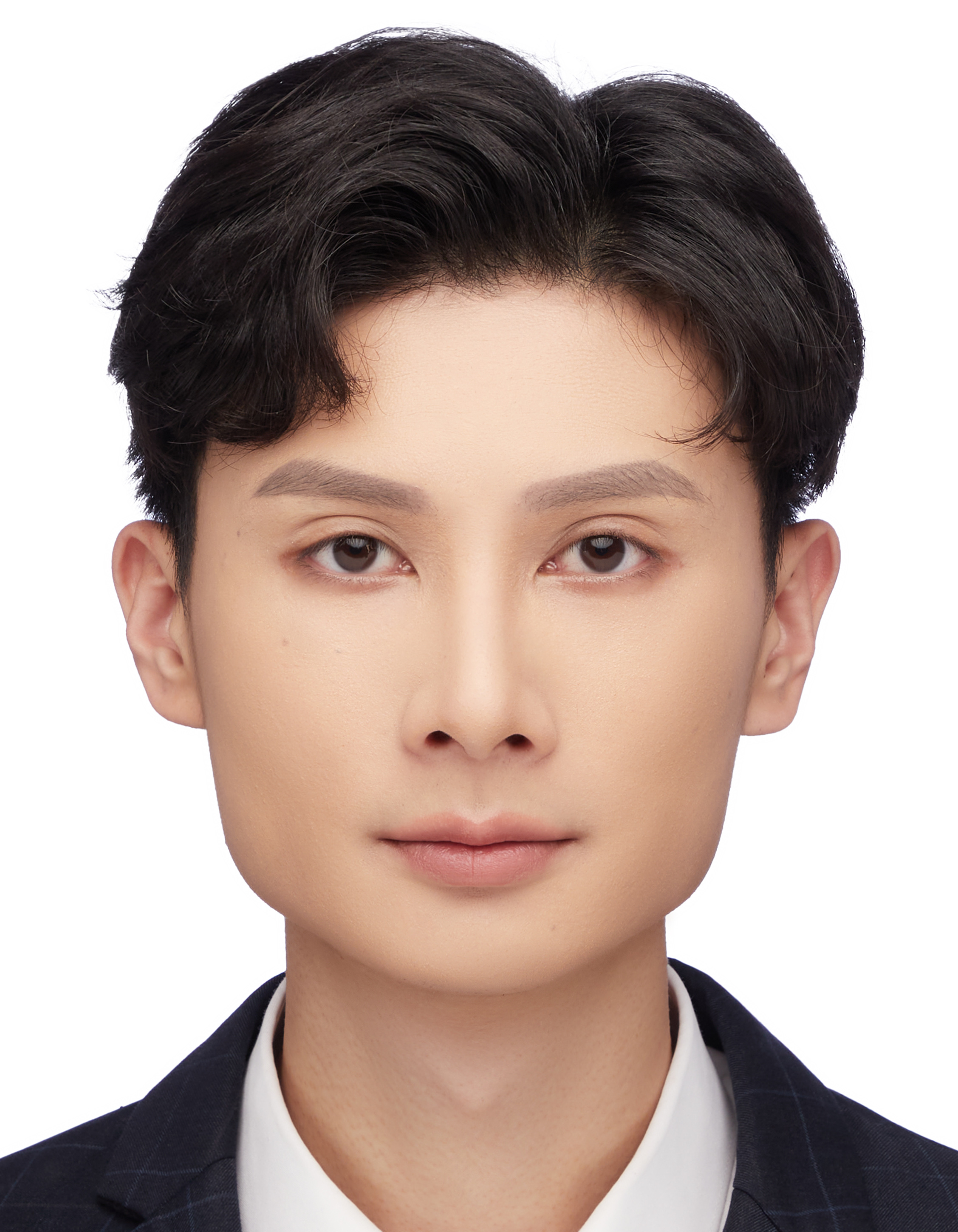}}]{Chun-An Lin}
(Student Member, IEEE) is currently a Ph.D. Candidate in the Department of Electrical Engineering, National Chung Hsing University, Taiwan, and a Lecturer with the College of Computing and Informatics, Providence University, Taiwan. He received his M.S. degree from the Department of Multimedia Design, National Taichung University of Science and Technology, Taiwan, in 2021 and his B.S. degree from the Department of Computer Science \& Communication Engineering, Providence University, Taiwan, in 2019. His research focuses on Machine Learning, Deep Learning, Image Super-Resolution, and Medical Image Analysis.
\end{IEEEbiography}

\begin{IEEEbiography}[{\includegraphics[width=1in,height=1.25in,clip,keepaspectratio]{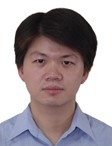}}]{Tsung-Jung Liu}
(Member, IEEE) received the B.S. degree in electrical engineering from National Tsing Hua University, Hsinchu, Taiwan, in 1998, the M.S. degree in communication engineering from National Taiwan University, Taipei, Taiwan, in 2001, and the Ph.D. degree in electrical engineering from the University of Southern California, Los Angeles, CA, USA, in 2014. He is currently an Associate Professor with the Department of Electrical Engineering, and the Graduate Institute of Communication Engineering, National Chung Hsing University, Taichung, Taiwan. His research interests include computer vision, perceptual image/video processing, visual quality assessment, image restoration, machine learning/deep learning, and artificial intelligence.
\end{IEEEbiography}

\begin{IEEEbiography}[{\includegraphics[width=1in,height=1.25in,clip,keepaspectratio]{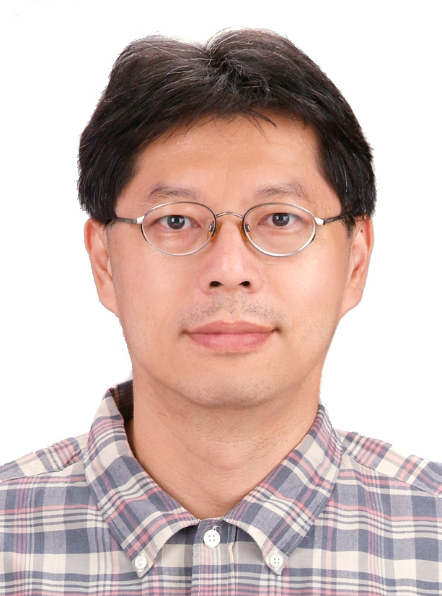}}]{Yen-Chieh Ouyang}
(Life Member, IEEE) received the Bachelor of Science degree in electrical engineering from Feng Chia University, Taiwan, in 1981, and the Master of Science and Doctor of Philosophy degrees in electrical engineering from The University of Memphis, Memphis, TN, USA, in 1987 and 1992, respectively. In August 1992, he joined the Department of Electrical Engineering, National Chung Hsing University, Taiwan, where he is currently a Professor. His research expertise lies in information-theoretical image processing with applications in medical imaging and remote sensing. He has significant experience in developing hyperspectral imaging models for magnetic resonance (MR) imaging applications and in the software engineering aspects of implementing and testing these methods. Furthermore, his research interests extend to the development and evaluation of cloud computing systems for various real-world applications. He also integrates artificial intelligence, machine learning, cognitive science, and deep neural networks into his work, applying these to intelligent systems such as decision support systems, human-computer interaction, and machine learning, as well as to algorithmic problems aimed at improving efficiency. His diverse research interests include: hyperspectral image processing, medical image processing, agricultural production applications, communication networks, network security in mobile networks, cloud computing, multimedia system design, and performance evaluation.
\end{IEEEbiography}

\end{document}